\documentclass[%
 reprint,
 amsmath,amssymb,
 aps,
]{revtex4-1}
\usepackage{multirow}
\usepackage{graphicx}
\usepackage{dcolumn}
\usepackage{booktabs}
\usepackage{makecell}
\usepackage{threeparttable}
\usepackage{bm}

\usepackage{siunitx}
\usepackage{hyperref}
\usepackage{footnote}
\usepackage{tikz}
\usepackage{tikz-feynman}
\usepackage{float}
\allowdisplaybreaks[4]

\newcommand{\bra}[1]{\langle #1 |}
\newcommand{\ket}[1]{| #1 \rangle}

\newcommand{\lra}[1]{\left(#1\right)}

\tikzfeynmanset{
every vertex/.style={red, dot},
every particle/.style={blue},
every blob/.style={draw=black!40!black, pattern color=black!40!black},
}
\usetikzlibrary{decorations.markings}
\usetikzlibrary{snakes}
\tikzset{
 photon/.style={decorate, decoration={snake}, draw=black},
    electron/.style={draw=black, postaction={decorate},
        decoration={markings,mark=at position .55 with {\arrow[draw=black]{>}}}},
    gluon/.style={decorate, draw=magenta,
        decoration={coil,amplitude=3pt, segment length=4pt}},
    scalar/.style={dashed,line width=.6pt, postaction={decorate},
        decoration={markings,mark=at position .55 with {\arrow[draw=black]{>}}}},
}

\begin{document}

\title{Prediction of an exotic state around 4240 MeV with $J^{PC}=1^{-+}$ \\
as the C-parity partner of Y(4260) in molecular picture}

\author{Xiang-Kun Dong$^{1,2}$}
 \ \email{dongxiangkun@itp.ac.cn}
 \author{Yong-Hui Lin$^{1,2}$}
 \ \email{linyonghui@itp.ac.cn}
\author{Bing-Song Zou$^{1,2,3}$}
\ \email{zoubs@itp.ac.cn}

\address{%
$^1$ CAS Key Laboratory of Theoretical Physics, Institute of Theoretical Physics,\\
Chinese Academy of Sciences, Beijing 100190, China\\
$^2$ University of Chinese Academy of Sciences, Beijing 100049, China\\
$^3$ School of Physics, Central South University, Changsha 410083, China
}


\begin{abstract}
The possibility of the Y(4260) being the molecular state of $D\bar D_1(2420)+\mathrm{c.c.}$ is investigated in the one boson exchange model. It turns out that the potential of $J^{PC}=1^{--}$ state formed by  $D\bar D_1(2420)+\mathrm{c.c.}$ is attractive and strong enough to bind them together when the momentum cutoff $\Lambda \gtrsim 1.4$ GeV. To produce the Y(4260) with correct binding energy, we need $\Lambda\approx 2.1$ GeV. Besides, $D\bar D_1(2420)+\mathrm{c.c.}$ can also form a state with exotic quantum numbers, $J^{PC}=1^{-+}$, and its potential similar with that of the $J^{PC}=1^{--}$ state. Therefore, an exotic state with mass around 4240 MeV (called $\eta_{c1}(4240)$) is predicted to exist. Our estimation of the mass of the $J^{PC}=1^{-+}$ state in charmonium region is in agreement with those predicted by the chiral quark model and the lattice QCD. The possible decay modes and their relative widths are estimated and the results suggest that this exotic state can be searched for in $\eta\eta_c$ and $\eta\chi_{c1}$ channels.
\end{abstract}

\pacs{Valid PACS appear here}
\maketitle


\section{Introduction}
In 2005 a vector charmoniumlike state, the $Y(4260)$, was reported by BaBar Collaboration~\cite{Aubert:2005rm}  in the initial-state radiation process $e^+e^-\to\gamma_{\mathrm {ISR}}J/\psi\pi^+\pi^-$ with a mass of $(4259\pm8^{+2}_{-6})$ MeV and a width of 50 $\sim$ 90 MeV, which was confirmed by CLEO Collaboration~\cite{He:2006kg} and Belle Collaboration~\cite{Yuan:2007sj} later. Recently, the $e^+e^-\to\pi^{+} \pi^{-} J / \psi$ cross section reported by BESIII~\cite{Ablikim:2016qzw} showed that the $Y(4260)$ contains two substructures, the $Y(4220)$ and $Y(4320)$, and the $Y(4220)$ is consistent with the previous $Y(4260)$. A combined analysis of BESIII data in four channels, $e^{+} e^{-} \rightarrow \omega \chi_{c 0}$~\cite{Ablikim:2015uix}, $\pi^{+} \pi^{-} h_{c}$~\cite{BESIII:2016adj}, $\pi^{+} \pi^{-} J / \psi$~\cite{Ablikim:2016qzw} and $ D^{0} D^{*-} \pi^{+}$ + c.c.~\cite{Ablikim:2018vxx}, yields a mass of $(4219.6 \pm 3.3 \pm 5.1)$ MeV and a width of $ (56.0 \pm 3.6 \pm 6.9)$ MeV~\cite{Gao:2017sqa}. Different analyses give quite different resonant parameters of $Y(4260)$ and the average values in the latest PDG~\cite{Tanabashi:2018oca} read $m=(4230\pm 8)$ MeV and $\Gamma=(55\pm19)$ MeV. 

It is clear that the $Y(4260)$ contains $c\bar c$ quarks and is above the thresholds for $D\bar D$, $D\bar D^*+\mathrm{c.c.}$ and $D^*\bar D^*$. However, no signals of the $Y(4260)$ appear in these channels~\cite{Abe:2006fj,Pakhlova:2008zza,Aubert:2008pa}, which indicates that it is not a conventional charmonium. Besides, there seems no room for the $Y(4260)$ in the $1^{--}$ $c\bar c$ spectrum~\cite{Godfrey:1985xj}. As a candidate for the exotic meson, its nature still remains controversial and has been attracting much attention. Several models were proposed to account for the peculiar behaviors of the $Y(4260)$ including a hybird state~\cite{Zhu:2005hp,Close:2005iz,Kalashnikova:2008qr}, an excited charmonium~\cite{LlanesEstrada:2005hz,Li:2009zu,Shah:2012js}, a baryonium composed of $\Lambda_c\bar\Lambda_c$~\cite{Qiao:2005av}, a hadrocharmonium~\cite{Dubynskiy:2008mq,Li:2013ssa}, a tetraquark state~\cite{Maiani:2005pe,Ali:2017wsf,Wang:2018ntv}, an interference effect~\cite{Chen:2010nv,Chen:2017uof} or a hadronic molecule of $\omega\chi_{c0}$~\cite{Dai:2012pb} or $\bar DD_1$ + c.c. See e.g. Ref.~\cite{Chen:2016qju} for a detailed discussion. 

Among
 these explanations, a molecular state of $D\bar D_1(2420)+\mathrm{c.c.}$ seems to be a good choice since the $Y(4260)$ is just below the threshold of $D\bar D_1(2420)+\mathrm{c.c.}$ and they can couple in S-wave. The mechanisms of the formation of the molecular $Y(4260)$ was discussed in Refs.~\cite{Ding:2008gr,Li:2013bca} by solving the Schr\"odinger equation with effective potential via meson exchange. The quenched lattice QCD also favors the $D\bar D_1(2420)+\mathrm{c.c.}$ molecule explanation~\cite{Chiu:2005ey}. Although it was argued in Ref.~\cite{Li:2013yka} that the production of $D\bar D_1(2420)+\mathrm{c.c.}$ in the electron-positron collisions is forbidden in the heavy quark limit due to the heavy quark spin symmetry (HQSS) and in turn suppressed in the real world, Ref.~\cite{Wang:2013kra} showed that the HQSS breaking is strong enough so that the molecule interpretation of the $Y(4260)$ does not contradict with the current experimental data. From the light-quark perspective, it is claimed that the $Y(4260)$ has a sizeable $D\bar D_1(2420)+\mathrm{c.c.}$ component, which is, however, not completely dominant~\cite{Chen:2019mgp}. By assuming the $Y(4260)$ being the $D\bar D_1(2420)+\mathrm{c.c.}$ molecule, its properties have been discussed in Refs.~\cite{Wang:2013kra,Li:2013yla,Wang:2013cya,Wu:2013onz,Qin:2016spb,Cleven:2013mka,Lu:2017yhl}. Furthermore, such interpretation is supported by the new experimental data: 
the observations of $Z_c(3900)\pi$~\cite{Ablikim:2013mio,Liu:2013dau} and $X(3872)\gamma$~\cite{Ablikim:2013dyn} in the mass region of the $Y(4260)$. We refer to Ref.~\cite{Guo:2017jvc} for more details of this molecular picture.

The $D\bar D_1(2420)+\mathrm{c.c.}$ can also form a system with positive C-parity, which is definitely an exotic state, if exists, since $J^{PC}=1^{-+}$ is not allowed for traditional $q\bar q$ mesons. Within the chiral quark model, Ref.~\cite{Li:2013bca} showed that the $D\bar D_1(2420)+\mathrm{c.c.}$ with $J^{PC}=1^{-+}$ can form a bound state with a mass of $4253\sim4285$ MeV. Besides, it is predicted by using the lattice QCD~\cite{Liu:2012ze} that the $J^{PC}=1^{-+}$ state in the charmonium region has a mass of $m(1^{-+})=m_{\eta_c}+1233\pm 16\ \mathrm{MeV}=4217\pm 16\ \mathrm{MeV}$, just below the Y(4260), which gives us more confidence in the existence of the $D\bar D_1(2420)+\mathrm{c.c.}$ bound state with $J^{PC}=1^{-+}$. On the other hand, the production and the decay of such exotic state were discussed in Ref.~\cite{Wang:2014wga} under the assumption of the Y(4260) being a molecule of $D\bar D_1(2420)+\mathrm{c.c.}$ where some guidance for the experiments was given.

In this paper we use the vector meson exchange interaction between $D\bar D_1(2420)+\mathrm{c.c.}$ to investigate whether it is possible for them to form the $J^{PC}=1^{--}$ and $J^{PC}=1^{-+}$ molecules. In addition, we also discuss the influence of $\sigma$ exchange on the potential. The possible bound states of $D\bar D_1(2430)+\mathrm{c.c.}$ via one boson exchange has also studied in Ref.~\cite{Shen:2010ky}. They showed that the $J^{PC}=1^{-+}$ system is attractive and can form a bound state while the $J^{PC}=1^{--}$ system is repulsive, which conflicts with the chiral quark model~\cite{Li:2013bca}. In Refs.~\cite{Close:2009ag,Close:2010wq}, Y(4260) were interpreted as a bound state of $D_1(2430)\bar D^*$ and its exotic C-parity partner was predicted in the vicinity of Y(4260). However, the binding energies in this case are more than one hundred MeV, which we think is too deep for a hadronic molecule. Such deep bound states were also disfavored in Ref.~\cite{Filin:2010se} since its signal disappear due to its too large width. Note that there are two $D_1$ states with similar masses while quite different decay widths~\cite{Tanabashi:2018oca}, 
\begin{align}
D_1(2420):\  m&=2420.8\pm0.5 \rm{MeV},\notag\\
 \Gamma&=31.7\pm2.5 \rm{MeV},\notag\\
D_1(2430):\ m&=2427\pm40 \rm{MeV}, \notag\\
\Gamma&=384^{+130}_{-110} \rm{MeV}.\notag 
\end{align}
We only use the narrow one (denoted by $D_1$ throughout the rest of the paper) since $D_1(2430)$ is too wide to form a molecular state~\cite{Filin:2010se}. We assume that the potential between the components of the $D\bar D_1+\mathrm{c.c.}$ molecule is dominated by the vector meson exchange interactions since the pseudoscalar meson exchange between $D\bar D_1+\mathrm{c.c.}$ is forbidden by parity conservation~\cite{Close:2008hv}. It is different from the assumption in Ref.~\cite{Ding:2008gr} where the Y(4260) was considered as the molecule of $DD_1$ or $D_0D^*$ through pseudoscalar mesons exchange (off-diagonal potential) and $\sigma$ exchange (diagonal potential). The vector mesons exchange was not included because some of the related coupling constants were not available. We emphasize that this picture is not advisable because $D_0$ is too wide to be the component of the Y(4260).

This paper is organized as follows: In section II, the meson exchange potential between $D\bar D_1+\mathrm{c.c.}$ is derived and the bound states for $D\bar D_1+\mathrm{c.c.}$ with both $J^{PC}=1^{--}$ and $1^{-+}$ are produced; The possible decay channels of the $1^{-\pm}$ molecules and their decay widths are discussed in Section III; We finally give a short summary and discussion in Section IV.

\section{Binding}
\subsection{C-Parity Conventions}
Both $D$ and $D_1$ are not the eigenstates of C-parity so the phases of $D$ and $D_1$ under charge conjugation transformation are not fixed. We adopt the following conventions
\begin{align}
\mathcal C\ket{D_{(1)}}=\ket{\bar D_{(1)}}\label{eq:C_Con_D1}
\end{align}
and the flavor wave functions of positive and negative C-parity $\ket{D\bar D_1}$ states now read
\begin{align}
C&=\pm:\frac{1}{\sqrt 2}\lra{\ket{D\bar D_1}\pm \ket{\bar DD_1}}
\end{align}
Actually, the physical results do not depend on C-parity conventions as long as the related Lagrangians are consistent with such conventions.

The potential between $D(\bar D)$ and $\bar D_1(D_1)$ is related to the corresponding scattering amplitude. For the state with $J^{PC}=1^{-\pm}$, the element of $S$ matrix reads
\begin{align}
&\ \ \ \ \bra{\bar DD_1\pm D\bar D_1}S\ket{\bar DD_1\pm D\bar D_1}\notag\\
&=\bra{\bar DD_1}S\ket{\bar DD_1}+\bra{D\bar D_1}S\ket{D\bar D_1}\notag\\
&\pm\left(\bra{\bar DD_1}S\ket{D\bar D_1}+\bra{D\bar D_1}S\ket{\bar DD_1}\right).
\end{align}

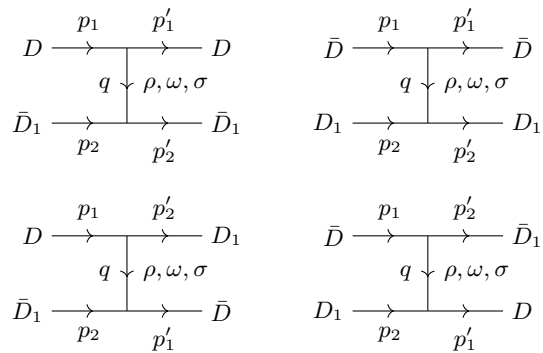
\begin{figure}
\centering
\begin{tikzpicture}
\draw[electron] (-1,0.5)--(0,0.5);
\draw[electron] (0,0.5)--(1,0.5);
\draw[electron] (-1,-0.5)--(0,-0.5);
\draw[electron] (0,-0.5)--(1,-0.5);
\draw[electron] (0,0.5)--(0,-0.5);
\node[right] at (0.1,0){$\rho,\omega,\sigma$};
\node[left] at (-1,0.5){$D$};
\node[left] at (-1,-0.5){$\bar D_1$};
\node[right] at (1,0.5){$D_1$};
\node[right] at (1,-0.5){$\bar D$};
\node[left] at (-0.1,0){$q$};
\node [above] at (-0.5,0.6){$p_1$};
\node [above] at (0.5,0.6){$p'_2$};
\node [below] at (-0.5,-0.6){$p_2$};
\node [below] at (0.5,-0.6){$p'_1$};

\begin{scope}[xshift=4cm]
\draw[electron] (-1,0.5)--(0,0.5);
\draw[electron] (0,0.5)--(1,0.5);
\draw[electron] (-1,-0.5)--(0,-0.5);
\draw[electron] (0,-0.5)--(1,-0.5);
\draw[electron] (0,0.5)--(0,-0.5);
\node[right] at (0.1,0){$\rho,\omega,\sigma$};
\node[left] at (-1,0.5){$ \bar D$};
\node[left] at (-1,-0.5){$ D_1$};
\node[right] at (1,0.5){$\bar D_1$};
\node[right] at (1,-0.5){$ D$};
\node[left] at (-0.1,0){$q$};
\node [above] at (-0.5,0.6){$p_1$};
\node [above] at (0.5,0.6){$p'_2$};
\node [below] at (-0.5,-0.6){$p_2$};
\node [below] at (0.5,-0.6){$p'_1$};
\end{scope}

\begin{scope}[yshift=2.5cm]

\draw[electron] (-1,0.5)--(0,0.5);
\draw[electron] (0,0.5)--(1,0.5);
\draw[electron] (-1,-0.5)--(0,-0.5);
\draw[electron] (0,-0.5)--(1,-0.5);
\draw[electron] (0,0.5)--(0,-0.5);
\node[right] at (0.1,0){$\rho,\omega,\sigma$};
\node[left] at (-1,0.5){$D$};
\node[left] at (-1,-0.5){$\bar D_1$};
\node[right] at (1,0.5){$D$};
\node[right] at (1,-0.5){$\bar D_1$};
\node[left] at (-0.1,0){$q$};
\node [above] at (-0.5,0.6){$p_1$};
\node [above] at (0.5,0.6){$p'_1$};
\node [below] at (-0.5,-0.6){$p_2$};
\node [below] at (0.5,-0.6){$p'_2$};
\end{scope}

\begin{scope}[xshift=4cm,yshift=2.5cm]

\draw[electron] (-1,0.5)--(0,0.5);
\draw[electron] (0,0.5)--(1,0.5);
\draw[electron] (-1,-0.5)--(0,-0.5);
\draw[electron] (0,-0.5)--(1,-0.5);
\draw[electron] (0,0.5)--(0,-0.5);
\node[right] at (0.1,0){$\rho,\omega,\sigma$};
\node[left] at (-1,0.5){$ \bar D$};
\node[left] at (-1,-0.5){$ D_1$};
\node[right] at (1,0.5){$\bar D$};
\node[right] at (1,-0.5){$ D_1$};
\node[left] at (-0.1,0){$q$};
\node [above] at (-0.5,0.6){$p_1$};
\node [above] at (0.5,0.6){$p'_1$};
\node [below] at (-0.5,-0.6){$p_2$};
\node [below] at (0.5,-0.6){$p'_2$};
\end{scope}
\end{tikzpicture}

\caption{Feynman diagrams for vector meson exchange between $D\bar D_1+\mathrm{c.c.}$. The top/bottom two are called direct/cross processes in this work.  The cross processes have opposite signs between $C=+$ and $C=-$ cases.}\label{fig:Feyn_DD1}
\end{figure}

There are four Feynman diagrams for $D\bar D_1+\mathrm{c.c.}$ elastic scattering by one boson (vector mesons and $\sigma$) exchange, shown in Fig.~(\ref{fig:Feyn_DD1}). Note that the scattering amplitudes of cross processes (bottom two in Fig.~(\ref{fig:Feyn_DD1})) in the positive and negative C-parity cases carry opposite signs and in turn yield opposite potentials.

\subsection{The vector exchange potential}
\subsubsection{The Lagrangian}
The couplings of heavy mesons and light vector meson nonet can be described by the effective Lagrangians, which satisfies the hidden gauge symmetry \cite{Casalbuoni:1996pg}. For $D$ and $D_1$ mesons, the Lagrangians read explicitly \cite{Ding:2008gr} 

\begin{align}
\mathcal{L}_{\mathrm{DDV}}&=i g_{\mathrm{DDV}}\left(\mathrm{D}_{b} \stackrel{\leftrightarrow}{\partial}_{\mu} \mathrm{D}_{a}^{\dagger}\right) V_{b a}^{\mu}\notag\\
&+i g_{\mathrm{\bar D} \mathrm{\bar D}\mathrm V} \left(\overline{\mathrm{D}}_{b} \stackrel{\leftrightarrow}{\partial}_{\mu} \overline{\mathrm{D}}_{a}^{\dagger}\right) V_{a b}^{\mu}\label{eq:LDDV}\\
\mathcal{L}_{\mathrm{D}_{1} \mathrm{D}_{1} V}&=i g_{\mathrm{D}_{1} \mathrm{D}_{1} V}\left(\mathrm{D}_{1 b}^{\nu} \stackrel{\leftrightarrow}{\partial}_{\mu} \mathrm{D}_{1 a \nu}^{\dagger}\right) V_{b a}^{\mu}\notag\\
&+i g'_{\mathrm{D}_{1} \mathrm{D}_{1} V}\left(\mathrm{D}_{1 b}^{\mu} \mathrm{D}_{1 a}^{\nu \dagger}-\mathrm{D}_{1 a}^{\mu \dagger} \mathrm{D}_{1 b}^{\nu}\right)\left(\partial_{\mu} V_{\nu}-\partial_{\nu} V_{\mu}\right)_{b a}\notag\\
&+i g_{\overline{\mathrm{D}}_{1}\overline{\mathrm{D}}_{1} V}\left(\overline{\mathrm{D}}_{1 b \nu} \stackrel{\leftrightarrow}{\partial}_{\mu} \overline{\mathrm{D}}_{1 a}^{\nu \dagger}\right) V_{a b}^{\mu}\notag\\
&+i g'_{\overline{\mathrm{D}}_{1}\overline{\mathrm{D}}_{1} V}\left(\overline{\mathrm{D}}_{1 b}^{\mu} \overline{\mathrm{D}}_{1 a}^{\nu \dagger}-\overline{\mathrm{D}}_{1 a}^{\mu \dagger} \overline{D}_{1 b}^{\nu}\right)\left(\partial_{\mu} V_{\nu}-\partial_{\nu} V_{\mu}\right)_{a b}\label{eq:LD1D1V}\\
\mathcal{L}_{\mathrm{DD}_{1} V}&=g_{\mathrm{DD}_{1} V} \mathrm{D}_{1 b}^{\mu} V_{\mu b a} \mathrm{D}_{a}^{\dagger}\notag\\
&+g_{\mathrm{DD}_{1} V}^{\prime}\left(\mathrm{D}_{1 b}^{\mu} \stackrel{\leftrightarrow}{\ \partial^{\nu}} \mathrm{D}_{a}^{\dagger}\right)\left(\partial_{\mu} V_{\nu}-\partial_{\nu} V_{\mu}\right)_{b a}\notag\\
&+g_{\overline{\mathrm{D}} \overline{\mathrm{D}}_{1} V}\overline{\mathrm{D}}_{a}^{\dagger} V_{\mu a b} \overline{D}_{1 b}^{\mu}\notag\\
&+g_{\overline{\mathrm{D}} \overline{\mathrm{D}}_{1} V}^{\prime}\left(\overline{\mathrm{D}}_{1 b}^{\mu} \stackrel{\leftrightarrow}{\ \partial^{\nu}} \overline{\mathrm{D}}_{a}^{\dagger}\right)\left(\partial_{\mu} V_{\nu}-\partial_{\nu} V_{\mu}\right)_{a b}+h . c .\label{eq:LDD1V}
\end{align}
where
\begin{align}
D_{(1)}&=(D_{(1)}^0,D_{(1)}^+)\\
V&=\left(\begin{array}{cc}{\frac{\rho^{0}}{\sqrt{2}}+\frac{\omega}{\sqrt{2}}} & {\rho^{+}}  \\ {\rho^{-}} & {-\frac{\rho^{0}}{\sqrt{2}}+\frac{\omega}{\sqrt{2}}}  \end{array}\right)
\end{align}
and
\begin{align}
g_{\mathrm{DDV}}&=-g_{\mathrm{\bar D} \mathrm{\bar D}\mathrm V}=\frac{1}{\sqrt{2}} \beta g_{V}\label{eq:gDDV}\\
g_{\mathrm{D_1D_1V}}&=-g_{\mathrm{\bar D_1} \mathrm{\bar D_1}\mathrm V}=\frac{1}{\sqrt{2}} \beta_2 g_{V}\\
g'_{\mathrm{D_1D_1V}}&=-g'_{\mathrm{\bar D_1} \mathrm{\bar D_1}\mathrm V}=\frac{5\lambda_2g_{V}}{3\sqrt{2}} M_{\rm D_1} \\
g_{\mathrm{DD_1V}}&=-g_{\mathrm{\bar D} \mathrm{\bar D_1}\mathrm V}=-\frac{2}{\sqrt{3}} \zeta_{1} g_{V} \sqrt{{M}_{\mathrm{D}} {M}_{\mathrm{D}_{1}}}\label{eq:gDD1V}\\
g'_{\mathrm{DD_1V}}&=-g'_{\mathrm{\bar D} \mathrm{\bar D_1}\mathrm V}=-\frac{1}{\sqrt{3}} \mu_1g_V.
\end{align}

It can be easily verified that the charge conjugation invariance of the above Lagrangians, Eqs.~(\ref{eq:LDDV}, \ref{eq:LD1D1V}, \ref{eq:LDD1V}), is consistent with the conventions, Eq.~(\ref{eq:C_Con_D1}).

\subsubsection{Estimation of coupling constants}

There are several parameters in the effective Lagrangians introduced in the last subsection. The already known ones are collected in the following,
\begin{align}
g_V&\approx 5.8,\\
\beta&\approx 0.9,\\
\lambda_1&\approx 0.1\ \mathrm{GeV}^2,
\end{align}
see Refs.~\cite{Bando:1987br}, \cite{Isola:2003fh} and \cite{Casalbuoni:1996pg}, respectively. These lead to $g_{\rm{DDV}}\approx 3.7$. The rest constants $\beta_2$, $\mu_1$ and $\zeta_1$ are not available now.

The $\mathcal L_{\rm{D}_1\rm{D}_1\rm{V}}$ contains two types of interaction, which are denoted by $g_{\mathrm{D_1D_1V}}$ and $g'_{\mathrm{D_1D_1V}}$ in Eq.~(\ref{eq:LD1D1V}). The second type vanishes in the nonrelativistic limit since $\partial_\mu V_\nu\sim q_\mu V_\nu$ and the exchanged four-momentum $q_\mu=(0,\bm{q})$ vanishes. Therefore, we only consider the first interaction, which has nothing to do with the angular momentum of $D_1$. As a rough estimation, we take $g_{\mathrm{D_1D_1V}}\approx -g_{\mathrm{DDV}}$ since they all describe the P-wave coupling of heavy mesons and the light vector meson. $\rm D$ and $\rm D_1$ have the same behaviors in such case where the spin of $D_1$ does not participate in. On the quark level, the force between isoscalar  light quark and antiquark is attractive by vector exchange~\cite{Zhang:2006ix,Li:2013bca}, so is the force between $\rm D$ and $\rm D_1$. Therefore, we adopt the minus sign here so that the vector exchange between $D$ and $\bar D_1$ leads to an attractive force, Eq.~(\ref{eq:V1r}). 

The $\mathcal L_{\rm{DD_1V}}$ also contains two types of interaction, denoted by $g_{\mathrm{DD_1V}}$ and $g'_{\mathrm{DD_1V}}$ in Eq.~(\ref{eq:LDD1V}). The first one describes the S-wave coupling, which dominates the interaction and hence the second one is neglected. In principle, the coupling constant $\xi_1$ in Eq.~(\ref{eq:gDD1V}) should be fixed from experimental data relating the $DD_1V$ vertex, for example the decay $D_1\to D\rho/\omega$. Unfortunately, such decay is forbidden kinetically.
  We notice that the coupling of $\rm{KK_1V}$ should be approximately the same as that of $\rm {DD_1V}$ because $s$ quarks in $\rm K$ and $\rm K_1$ and $c$ quarks in $\rm D$ and $\rm D_1$ are all spectators during the interactions. This treatment is supported by the following observations: 
\begin{itemize}
\item[a)]$g_{\rm{DDV}}=\beta g_{\rm{KKV}}\approx0.9g_{\rm{KKV}}$, see Eq.~(\ref{eq:gDDV}) and e.g. Ref~\cite{Zhang:2006ix},
\item[b)]$\frac{g_{\rm {D}^*\rm{D}\pi}}{\sqrt{m_{\rm{D}^*}m_{\rm{D}}}}\approx0.96 \frac{g_{\rm{K}^*\rm{K}\pi}}{\sqrt{m_{\rm{K}^*}m_{\rm{K}}}}$ in Refs.~\cite{Lin:2017mtz,Lin:2019qiv},
\item[c)]$g_{\rm{D}^*\rm{D}^*\rho}\approx0.84 g_{\rm{K}^*\rm{K}^*\rho}$ in Refs.~\cite{Lin:2019qiv,Lin:2018kcc}.
\end{itemize}
Therefore, we take 
\begin{equation}
\frac{g_{\rm{D}_1\rm{DV}}}{\sqrt{m_{\rm{D}_1}m_{\rm{D}}}}\approx0.9\frac{g_{\rm{K}_1\rm{KV}}}{\sqrt{m_{\rm{K}_1} m_{\rm K}}}\label{eq:gDK}
\end{equation}
as a rough estimation and $g_{\rm{K}_1\rm{KV}}$ can be fixed by the decay width of $\rm K_1\to \rm K\rho$. Finally we obtain
\begin{align}
g_{\rm{DD}_1\rm{V}}\approx 2.3\ \rm{GeV},
\end{align}
see Appendix \ref{sec:gdd1v} .

\subsubsection{The potential in position space}

For the vector exchange in the first two diagrams in Fig.~(\ref{fig:Feyn_DD1}), the scattering amplitude is
\begin{align}
i\mathcal M_1
= -ig_{\rm{DDV}}g_{\rm{ D_1 D_1V}}\frac{4m_{\rm D}m_{\rm{D}_1}}{|{\bf q}|^2+m_{\rm V}^2}
\end{align}
and the corresponding potential in momentum space reads
\begin{align}
\tilde V_{\rm v1}({\bf q},m_{\rm V})&=-\frac{\mathcal M_1}{4m_{\rm D}m_{\rm{D}_1}}\notag\\
&=g_{\rm{DDV}}g_{\rm{ D_1 D_1V}}\frac{1}{|{\bf q}|^2+m_{\rm V}^2}.
\end{align}
After Fourier transformation we obtain the potential in position space
\begin{align}
V_{\rm v1}({\bf r},m_V)&=g_{\rm{DDV}}g_{\rm{D_1 D_1V}}m_{\rm{V}}Y(m_{\rm{V}} r)\label{eq:V1r}
\end{align}
where $Y(x)=e^{-x}/4\pi x$ is the Yukawa potential.

For the vector exchange in the last two diagrams,
\begin{align}
i\mathcal M_2
\approx -ig^2_{\rm{DD_1V}}\lra{1+\frac{{\bm\epsilon}_1\cdot \bm{q\epsilon}_2\cdot {\bm q}}{m_{\rm{V}}^2}}\frac{1}{|{\bm q}|^2+\tilde{m}^2}
\end{align}
 where $\tilde m^2=m_{\rm{V}}^2-(m_{\rm{D}_1}-m_{\rm{D}})^2$.
$\epsilon_1$ and $\epsilon_2$ are the polarizations of initial and final $D_1$'s, respectively. The potential (see e.g. Refs.~\cite{Thomas:2008ja,Liu:2019stu} for more details of such Fourier transformation) reads
\begin{align}
V_{\rm v2}({\bf r},m_{\rm{V}})
&=\frac{g^2_{\rm{DD_1V}}}{4m_{\rm{D}}m_{{\rm{D}_1}}}\bigg[\lra{\tilde m -\frac{\tilde m^3}{3 m_{\rm{V}}^2}}Y(\tilde m r)\notag\\
&\ \ \ \ \ \ \ \ \ \ \ \ \ \ \ \ \ \ \ \ \left.+\frac{1}{3 m_{\rm{V}}^2}\delta^{(3)}(\bm r)\right].\label{eq:V2r}
\end{align}

Taking the isospin factor into account, we obtain
\begin{align}
V_{{\rm v}i}^{I=0}({\bf r},m_{\rm{V}})&=\frac12\lra{3V_{ i}({\bf r},m_\rho)+V_{ i}({\bf r},m_\omega)},\\
V_{{\rm v}i}^{I=1}({\bf r},m_{\rm{V}})&=\frac12\lra{V_{ i}({\bf r},m_\rho)-V_{ i}({\bf r},m_\omega)}.
\end{align}
where $i=1,2.$ For the $J^{PC}=1^{-\pm}$ state, the total vector exchange potential reads
\begin{align}
V^{C=\pm}_{\mathrm{v}}=V_{\rm v1}^{I}\pm V_{\rm v2}^{I}.\label{eq:potentialCpm}
\end{align}
where $I=0,1$. Note that $m_\rho\approx m_\omega$ so the potentials for isovector (I=1) are very weak. We only consider the isoscalar state here.
\subsubsection{Form factor}
Up to now we have treated every hadron as a point particle and the potentials diverge at the origin. To account for the finite size of actual hadrons and regularize the singularity of the potential, we introduce a monopole form factor
\begin{equation}
F(q,m,\Lambda)=\frac{\Lambda^2-m^2}{\Lambda^2-q^2}
\end{equation}
into each vertex with the same momentum cutoff $\Lambda$, which in position space can be looked upon as a spherical source of the exchanged mesons~\cite{Ericson:1982ei,Tornqvist:1993ng}. Actually, such form factor was initially introduced to describe the nucleon scattering via pion (or other light mesons) exchange (see e.g. Refs.~\cite{Holinde:1981se,Ericson:1982ei}) and now it is widely used in one boson exchange model (see e.g. Refs.~\cite{Ding:2008gr,Shen:2010ky,Zhang:2006ix}). The value of $\Lambda$ is model-dependent and varies in the range of $0.8\sim 2.0$ GeV from different analyses (see e.g. Ref.~\cite{Li:1996yn} and the discussion in Ref.~\cite{Holinde:1981se}). This is one of the origins of uncertainty in our model.

As discussed in Refs.~\cite{Eides:2017xnt,Liu:2019zvb,Burns:2019iih}, although the form factor makes the contact $\delta$ potential in Eq.~(\ref{eq:V2r}) finite, such term is still nonphysical because it introduces a much strong force in the short range, which is unexpected in the boson exchange model. Besides, such term may results in a repulsive force at long distance for an expected bound system, which is not self-consistent. Considering these arguments, we also exclude the $\delta$ potential in our calculation.

After introducing the form factor, the potentials in position space, Eqs.~(\ref{eq:V1r},\ref{eq:V2r}), are modified,
\begin{align}
V_{\rm v1}\left(\mathbf{r}, m_{{\rm{V}}}\right)&=g_{\rm{DDV}}g_{ \rm{ D_1 D_1V}}{\Big(}m_{\rm{V}}Y(m_{\rm{V}} r)\notag\\
&\ \ \ \ -\Lambda Y(\Lambda r)-\frac{1}{2}(\Lambda^{2}-m_{\rm{V}}^{2})r Y(\Lambda r)\Big),\\
V_{\rm v2}\left(\mathbf{r}, m_{{\rm{V}}}\right)&=\frac{g^2_{\rm{DD_1V}}}{4m_{\rm{D}}m_{{\rm{D}_1}}}\Big\{\Big[\tilde{m}_{\rm{V}}Y(\tilde{m}_{\rm{V}} r)-\tilde\Lambda Y(\tilde\Lambda r)\notag\\
&\ \ \ \ -\frac{1}{2}(\tilde\Lambda^{2}-\tilde m^{2}_{\rm{V}})r Y(\tilde\Lambda r)\Big]\notag\\
&\ \ \ \ -\frac{\tilde m^2_{\rm{V}}}{3m_{\rm{V}}^2}\Big[{\tilde{m}_{\rm{V}}Y(\tilde{m}_{\rm{V}} r)-\tilde\Lambda Y(\tilde\Lambda r)}\notag\\
&\ \ \ \ -\frac{1}{2}(\tilde\Lambda^{2}-\tilde m^{2}_{\rm{V}})r Y(\tilde\Lambda r)\Big]
\Big\}
\end{align}
with
\begin{align}
\tilde \Lambda^2&=\Lambda^2-(m_{\rm D_1}-m_{\rm D})^2.\label{eq:lambdatilde}
\end{align}
\subsection{The $\sigma$ exchange potential}
The $\sigma$ exchange potential has been deduced in Ref.~\cite{Ding:2008gr}. For the direct process,
\begin{align}
V_{\sigma1}\left(\mathbf{r}\right)&=-g_\sigma g_\sigma ''{\Big(}m_\sigma Y(m_\sigma r)-\Lambda Y(\Lambda r)-\notag\\
&\ \ \ \ \frac{1}{2}(\Lambda^{2}-m_\sigma^{2})r Y(\Lambda r)\Big)
\end{align}
and for the cross process,
\begin{align}
V_{\sigma2}&=\frac29\frac{h_\sigma'^2}{f_\pi^2}\tilde m^2\Big[{ \tilde m Y(\tilde m r)-\tilde \Lambda Y(\tilde\Lambda r)}\notag\\
&\ \ \ \ -\frac{1}{2 }(\tilde\Lambda^{2}-\tilde m^{2})r Y(\tilde\Lambda r)\Big]
\end{align}
with $\tilde m^2=m_\sigma^2-(m_{\rm D_1}-m_{\rm{D}})^2$ and $\tilde \Lambda$ the same as Eq.~(\ref{eq:lambdatilde}).
Here we have also excluded the $\delta$ potential. In our calculation, the constants in the above potentials are taken to be
\begin{align}
g_\sigma g_\sigma''&= 0.58,\\
h_\sigma'&=0.35,\\
f_\pi&= 132\ \rm{MeV}
\end{align}
as in Ref.~\cite{Ding:2008gr}.

The isospin factor is trivial in this case. For the $J^{PC}=1^{-\pm}$ state, the total $\sigma$ exchange potential reads
\begin{equation}
V^{C=\pm}_{\sigma}=V_{\sigma1}\pm V_{\sigma2}.
\end{equation}

\subsection{Binding Energies}

\begin{figure}[h]

\centering
\includegraphics[width=0.8\linewidth]{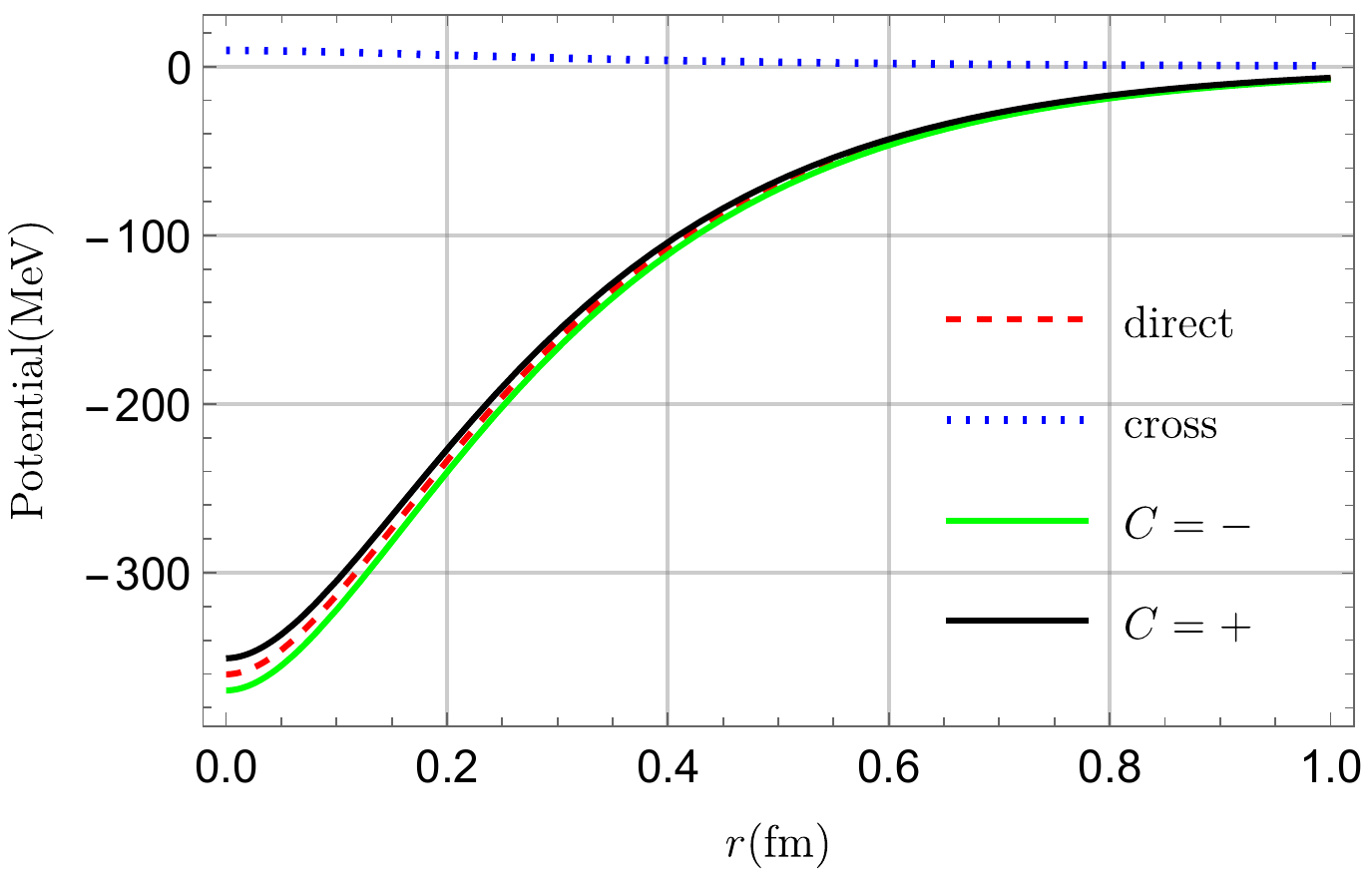}
\includegraphics[width=0.8\linewidth]{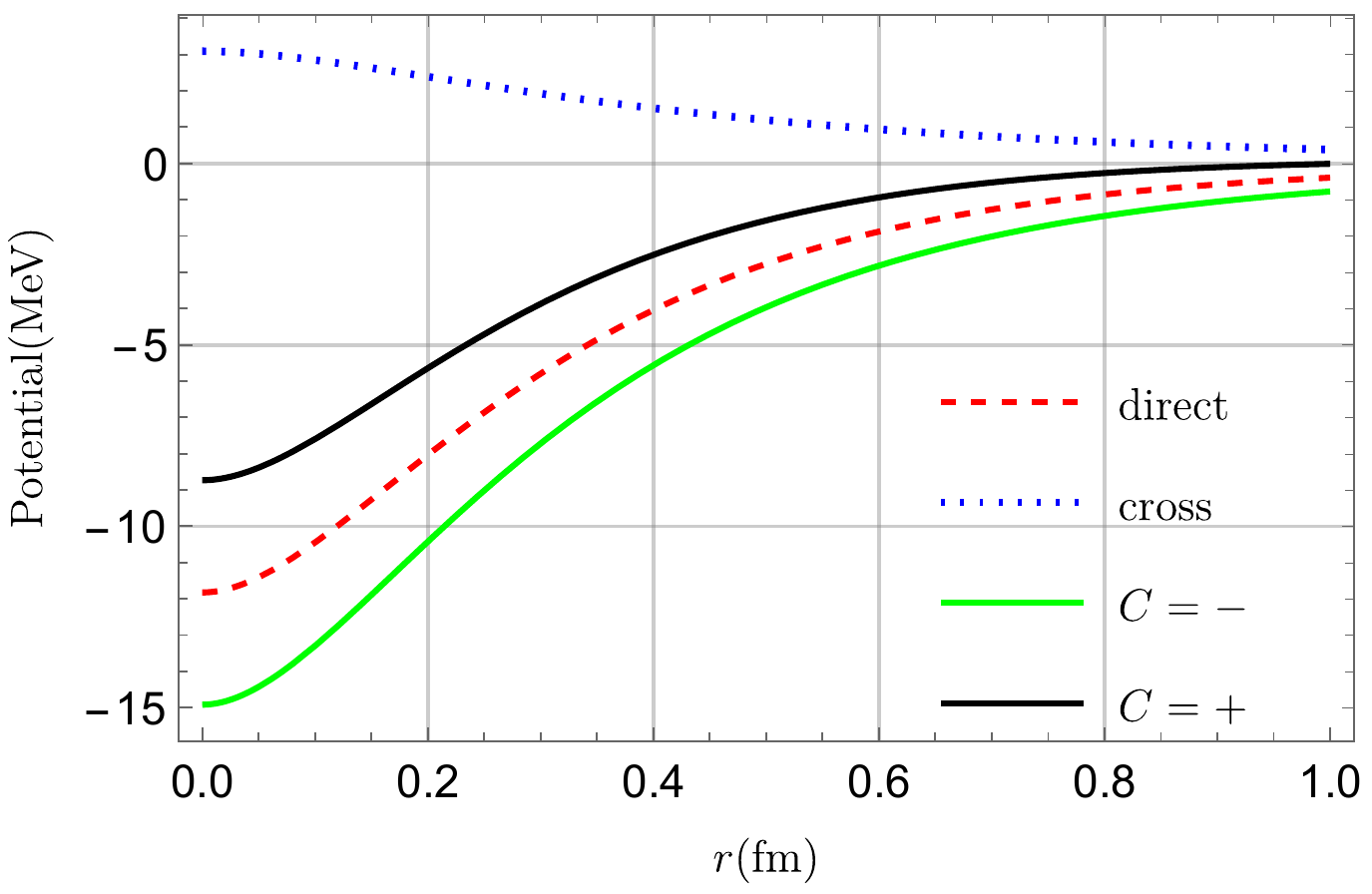}
\includegraphics[width=0.8\linewidth]{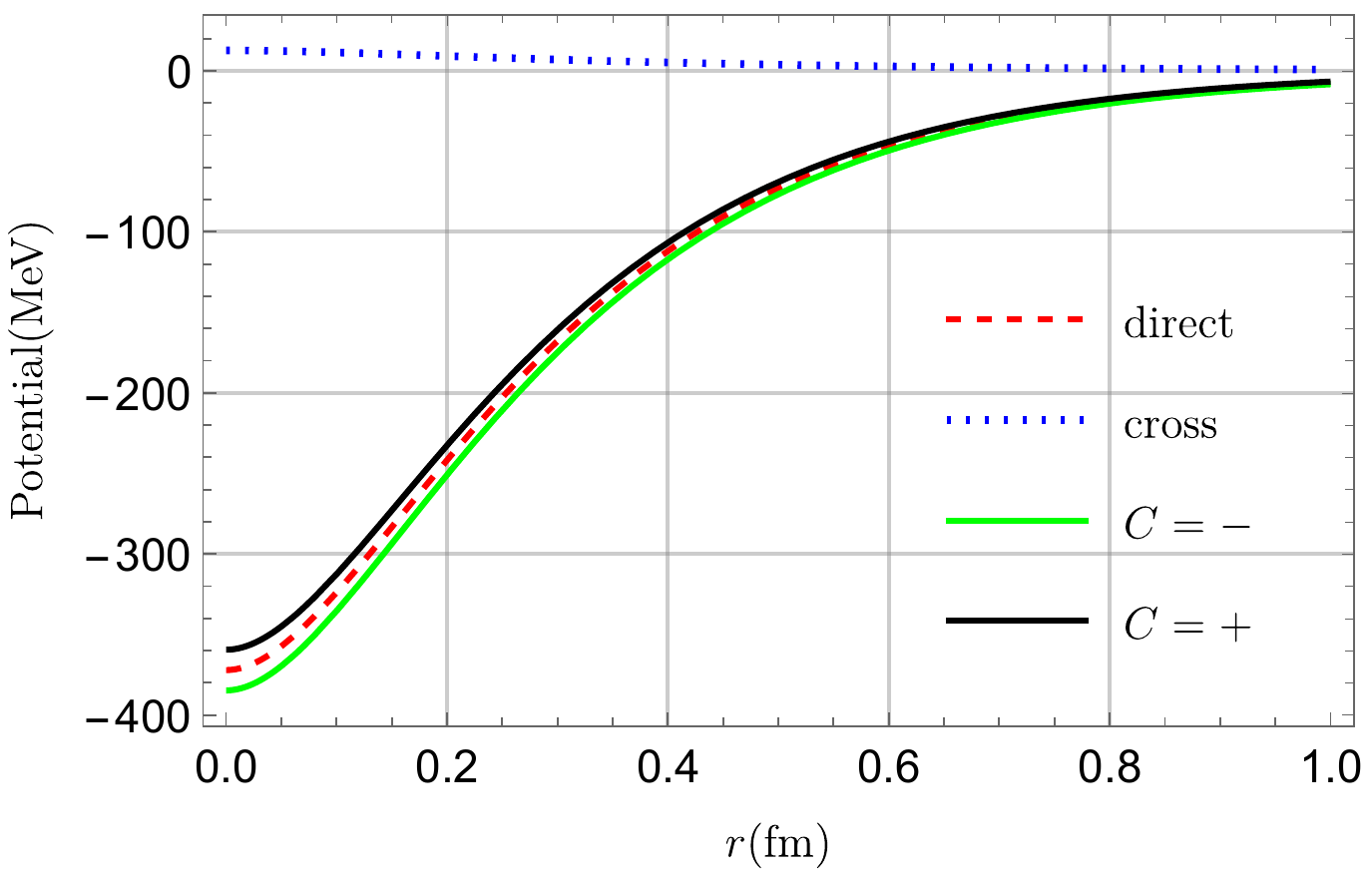}
\caption{The vector (top) or $\sigma$ (middle) exchange potentials and the total potentials (bottom) with $\Lambda=1.5$ GeV. The ``direct" represents the potential for the first two diagrams in Fig.~(\ref{fig:Feyn_DD1}) and the ``cross" for the second two diagrams. ``$C=+$" and ``$C=-$" represent the potentials for positive and negative C-parity states, respectively.}\label{fig:diff_poten}
\end{figure}

We use the following values from PDG \cite{Tanabashi:2018oca},
\begin{align}
m_{\rm{D}}&=1.867\ \rm{GeV},\\
m_{\rm{D_1}}&=2.420\ \rm{GeV},\\
m_\rho&=0.775\ \rm{GeV},\\
m_{\omega}&=0.783\ \rm{GeV},\\
m_{\sigma}&\approx 0.600 \ \rm{GeV}.
\end{align}
Besides, we take $g_{\rm{D_1DV}}\approx -g_{\rm{DDV}}\approx -3.7$, as analyzed above. Using the decay of $K_1$ we estimate $g_{\rm{D_1DV}}$ to be around 2.3 GeV. The vector and $\sigma$ exchange potentials  and the total potentials are shown in Fig.~\ref{fig:diff_poten} from where we see that
\begin{equation}
V^{C=+}_{\rm {Tot}}<V^{C=-}_{\rm {Tot}}<0.
\end{equation}
Therefore, it is expected that the $D\bar D_1$ bound state with $J^{PC}=1^{-+}$ should exist if the Y$(4260)$ can be interpreted as a $D\bar D_1$ bound state with $J^{PC}=1^{--}$.

\begin{figure}
\centering
\includegraphics[width=0.8\linewidth]{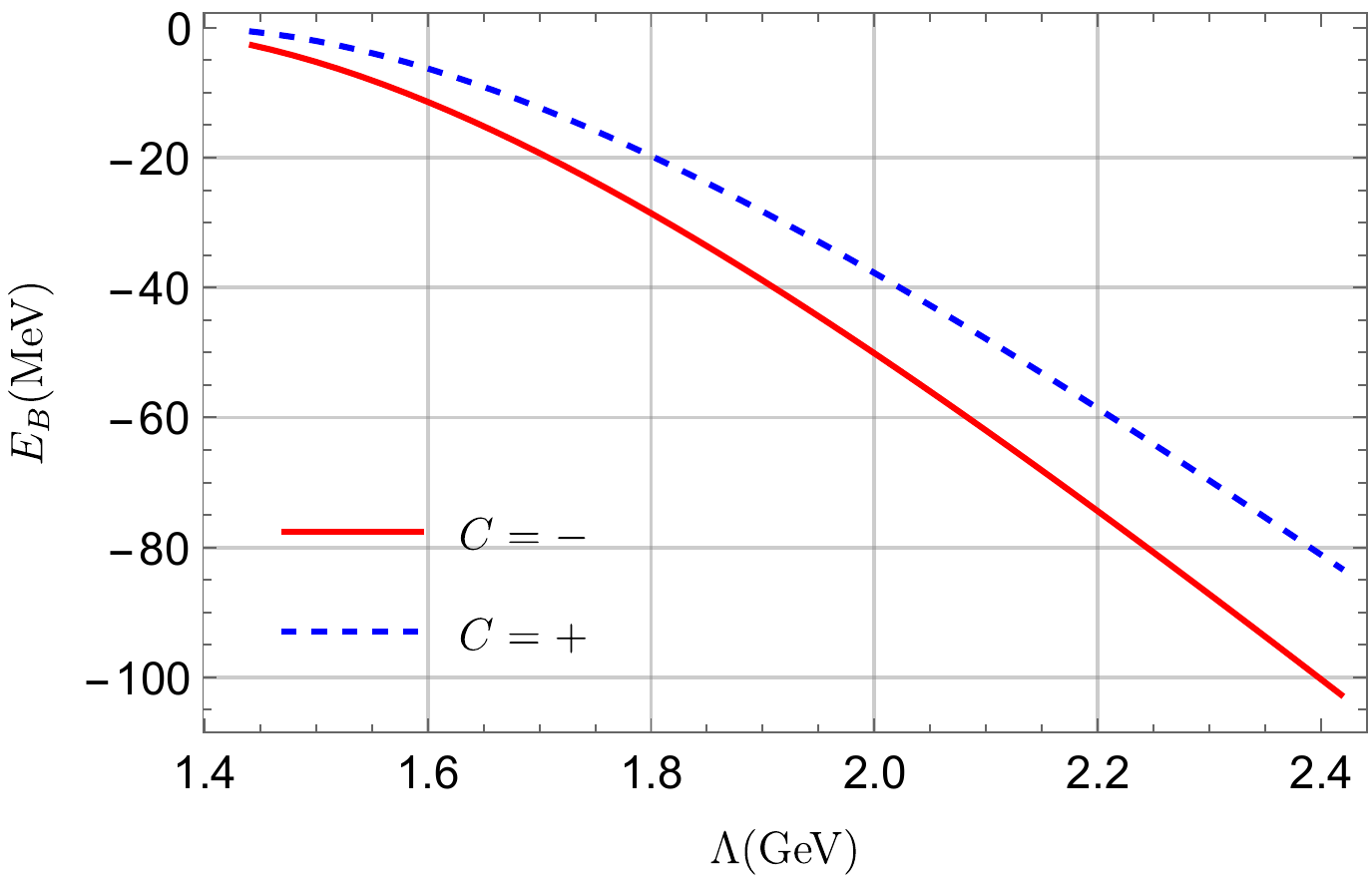}
\caption{Dependence of binding energies on the cutoff $\Lambda$.}\label{fig:bindingEs}
\end{figure}

The Schr\"odinger equations  for both $V^{C=+}_{\mathrm{v}}$ and $V^{C=-}_{\mathrm{v}}$ are solved and the dependence of binding energy on $\Lambda$ is shown in Fig.~\ref{fig:bindingEs}. Bound states comes into existence when $\Lambda \gtrsim 1.4$ GeV. Y(4260), as a pure bound state of $D\bar D_1$ with $J^{PC}=1^{--}$, has a binding energy of around 60 MeV and consequently we need a $\Lambda_0$ of around 2.1 GeV to produce such binding energy. In turn, this specific $\Lambda_0$ leads to a bound state of $J^{PC}=1^{-+}$ $D\bar D_1$ with a binding energy of about 70 MeV. We also include the $\sigma$ exchange potential and it turns out to be insignificant. If we assume that the Y(4260) is a pure $D\bar D_1+\mathrm{c.c.}$ bound state with $\Lambda_0\approx 2.1$ GeV, its $1^{-+}$ partner should has a mass around 4230 MeV. Generally, we do not expect a hadronic molecule to have such a big binding energy, namely 60 or 70 MeV. Since the Y(4260) may be a mixture of $D\bar D_1+\mathrm{c.c.}$ molecule and $\psi(nD)$ \cite{Lu:2017yhl}, the binding energy of $D\bar D_1+\mathrm{c.c.}$ molecule in Y(4260) is not necessarily so large. Confining $\Lambda$ in the commonly used range $0.8\sim 2.0$ GeV, we conclude that the $1^{-+}$ $D\bar D_1+\mathrm{c.c.}$ molecule has a binding energy $\lesssim 40$ MeV. Therefore, we expected this exotic $D\bar D_1+\mathrm{c.c.}$ molecule (called $\eta_{c1}(4240)$) to be around 4240 MeV. We should keep in mind that there are several uncertainties in our calculation from such as the coupling constants, the masses of involved hadrons, the form factor and so on, which are not take into account. It reminds us that we should not take the certain value $4240$ MeV seriously. Instead, what we can assert is that if we assume that there is $1^{--}$ $D\bar D_1+\mathrm{c.c.}$ molecule component in Y(4260), i.e., their attractive force is strong enough to bind themselves together, the $1^{-+}$ $D\bar D_1+\mathrm{c.c.}$ system is also at the edge of having a bound state.

\section{Decays}
With the masses of $D\bar D_1+\mathrm{c.c.}$ molecule states obtained, we can estimate their decay patterns by using the Effective Lagrangian Method as we did in the meson-baryon molecules sector previously~\cite{Lin:2017mtz,Lin:2018kcc,Lin:2019qiv}. The three-body $D\bar{D}^*\pi$ decay through the component meson $\bar{D}_1$ decaying directly into $\bar{D}^*\pi$ is considered naturally. All other possible two-body decay channels are collected in Table.~\ref{tab:channels}.
\begin{figure}
	\begin{minipage}{0.4\linewidth}
		\centering
	\includegraphics[width=\linewidth]{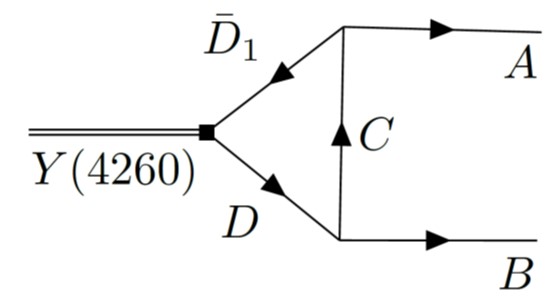}
	\end{minipage}
	\begin{minipage}{0.1\linewidth}
		
	\end{minipage}
	\begin{minipage}{0.45\linewidth}
		\centering
	\includegraphics[width=\linewidth]{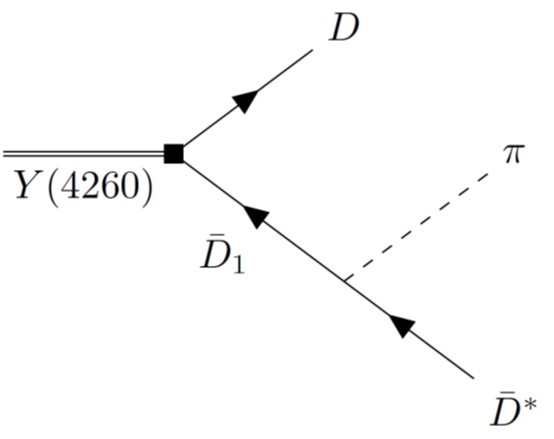}
	\end{minipage}
	
	\caption{Feynman diagrams for the 2- and 3-body decays of $Y(4260)$. They are similar for $\eta_{c1}(4240)$.}\label{fig:Feyn_2_3_decay}
\end{figure}
\begin{table}[htpb]
	\centering
	\caption{\label{tab:channels}Two-body decay channels for the $D\bar D_1+\mathrm{c.c.}$ molecule states considered in our calculation.}
	\begin{threeparttable}
		\scalebox{1}{
			\begin{tabular}{c|c|*{2}{c}}
				\Xhline{1.0pt}
				Molecule & Components & Final states & Exchanged particles \\
				\Xhline{0.8pt}
				\multirow{5}*{\thead{$1^{--}$\\~$Y(4260)$}}& \multirow{5}*{\thead{$D\bar D_1$\\  $+\mathrm{c.c.}$}} & $\omega\sigma$, $J/\psi \sigma$,$\omega\chi_{c0}$ & $D$ \\
				\Xcline{3-4}{0.4pt}
				& & $\bar D D^*$, $J/\psi \eta$, $\rho\pi$ & $D^*$, $\rho$, $\omega$ \\
				\Xcline{3-4}{0.4pt}
				& &  $h_c \eta$, $Z_c \pi$ & $D^*$ \\
				\Xcline{3-4}{0.4pt}
				& & $\bar D^* D^*$ & $\pi$ \\
				\Xcline{3-4}{0.4pt}
				& &  $\bar D D$ & $\rho$, $\omega$ \\
				\Xhline{0.8pt}
				\multirow{5}*{\thead{$1^{-+}$\\~$\eta_{c1}(4240)$} }&\multirow{5}*{\thead{$D\bar D_1$\\  $+\mathrm{c.c.}$}} & $\bar D D^*$, $\rho\pi$ & $D^*$, $\rho$, $\omega$ \\
				\Xcline{3-4}{0.4pt}
				& & $\chi_{c1}\eta$,  $Z_c \pi$ & $D^*$ \\
				\Xcline{3-4}{0.4pt}
				& & $\bar D^* D^*$, $J/\psi\omega$ & $\pi$, $D$ \\
				\Xcline{3-4}{0.4pt}
				& & $\eta_c\eta$ & $D^*$, $\rho$, $\omega$ \\
				\Xcline{3-4}{0.4pt}
				& & $\chi_{c0} \sigma$ & $D$ \\
				\Xhline{1.0pt}
			\end{tabular}
		}
	\end{threeparttable}
\end{table}
By convention, as illustrated in Fig.~(\ref{fig:Feyn_2_3_decay}), the three-body decay happens in the tree level and two-body decays happen through the triangle mechanism where a Gaussian regulator (at the vertex of $YD\bar D_1$) and a multipole form factor (in the propagator of the exchanged particle $C$), formulated as following, are included to study out the ultraviolet divergence.
\begin{align}
&f_1(p^2_E /\Lambda_0^2) = {\rm{exp}}(-p^2_E /\Lambda_0^2),\label{eq:gaussian}\\
&f_2(q^2) = \frac{\Lambda_1^4}{(m^2 - q^2)^2 + \Lambda_1^4}, \label{eq:multipolar}
\end{align}
where $p_E$ is the four dimensional Euclidean Jacobi momentum defined as $(m_1p_2-m_2p_1)/(m_1+m_2)$ with $m_1~(m_2)$ and $p_1~(p_2)$ the mass and momentum of $D~(\bar D_1)$, $m$ and $q$ is the mass and momentum of the exchanged particle. In addition to the $DD_1V$ vertex shown in Eq.~\eqref{eq:LDD1V}, the other effective Lagrangians that $D_1$ has involved in our calculation are listed as following,
\begin{align}
\mathcal{L}_{YD\bar{D}_1}&=g_{YD\bar{D}_1}\bar{D}_1^{\dagger \mu}Y_\mu D+h.c.,\\
\mathcal{L}_{D^*D_1P}&=g_{D^*D_1P}\Big[3D^{*\dagger\mu}D_1^\nu\partial_\mu\partial_\nu P-D^{*\dagger\mu}D_{1\mu}\partial^\nu\partial_\nu P\notag\\
&+\frac{1}{M_{D^*}M_{D_1}}\partial^\tau D^{*\dagger\mu}\partial^\nu D_{1\mu}\partial_\nu\partial_\tau P\Big]+h.c.,\label{eq:LDsD1P}\\
\mathcal{L}_{D^*D_1\pi}&=g_S D^{*\dagger\mu} D_{1\mu}\pi \notag\\
&+g_DD^{*\dagger\mu} D_{1}^\nu \Big(r_\mu r_\nu-\frac13 r^\beta r_\beta \tilde{g}_{\mu\nu}\Big)+h.c.,\label{eq:LDsD1pi}
\end{align}
with $\tilde{g}^{\mu\nu}\equiv\left(g^{\mu\nu}-\frac{p^\mu p^\nu}{M_{\rm D_1}^2}\right)$, $r^{\mu}\equiv\tilde{g}^{\mu\nu}\left(k_1-k_2\right)_\nu$. $p$, $k_1$ and $k_2$ are the momenta of $D_1$, $D^*$ and $\pi$, respectively. It should be mentioned that the Lagrangian $\mathcal{L}_{D^*D_1P}$ in Eq.~\eqref{eq:LDsD1P} can also be applied to the vertex $D_1D^*\pi$. $\mathcal{L}_{D^*D_1P}$ is a part of the heavy meson chiral perturbation~($\mathrm{HM\chi PT}$) theory which is constructed starting from the heavy quark spin-flavor and chiral symmetry and obtained by expanding the whole effective Lagrangian in $\mathrm{HM\chi PT}$ to the leading order of pseudo-Goldstone field. It implies the Lagrangian $\mathcal{L}_{D^*D_1P}$ in Eq.~\eqref{eq:LDsD1P} is only valid for the vertex where the momentum of the light boson can be treated as a relative small quantity. In the channels of $\pi\pi$ and $\rho\pi$, however, the magnitude of the three momentum of pion meson is around $2~\mathrm{GeV}$. Then another phenomenology interaction in Eq.~\eqref{eq:LDsD1pi} is used for the vertex $D_1D^*\pi$ in these channels and $g_S$, $g_D$ can be fitted to the various ratios of D-wave contribution in the process of $D_1$ decaying into $D^*\pi$ where the total width of $D_1$ is assumed to be saturated by the $D^*\pi$ completely.

The coupling constants between the molecule states and the component hadron pairs, $g_{YD\bar{D}_1}$, are estimated with the composite condition~\cite{Weinberg:1962hj,Weinberg:1965zz}. In the chiral and heavy quark limit, $g_{D^*D_1P}$ is defined as $-\sqrt{6} h^\prime /(3f_\pi)$ with $f_\pi=0.132~\mathrm{GeV}$ and $h^\prime=0.65~\mathrm{GeV}^{-1}$~\cite{Casalbuoni:1996pg}. And recent analysis on the $D_1D^*\pi$ coupling in Ref.~\cite{Guo:2020oqk} shows that the D-wave contribution is around half of the total width of $D_1$. This ratio leads to $g_S=2.52~\mathrm{GeV}$ and $g_D=10.6~\mathrm{GeV}^{-1}$ in Eq.~\eqref{eq:LDsD1pi}. All other conventional effective Lagrangian and couplings can be find in our previous work~\cite{Lin:2017mtz,Lin:2018kcc,Lin:2019qiv}. Note that cutoffs $\Lambda_0$ and $\Lambda_1$ are the free parameters in our calculation and we vary $\Lambda_1$ in the range of $1.5$-$2.4 \ \mathrm{GeV}$ to scrutinize how the decay behaviors undergo changes as the cutoffs are varied. $\Lambda_0$ is empirically adopted as $1~\mathrm{GeV}$ and the numerical results are shown in Table.~\ref{tab:widths}. In addition, Table.\ref{tab:widths1} shows the results when we get rid of the second form factor $f_2(q^2)$ in the loop integrals.

In the case of the form factor $f_2(q^2)$ kept and $\Lambda_0$ is taken as $1.0~\mathrm{GeV}$, it does not escape attention that $D^*\bar{D}^*$ is the dominant decay channel for both of $1^{--}$ and $1^{-+}$ $D\bar D_1+\mathrm{c.c.}$ molecules. The measured total width of $Y(4260)$ state, $\sim50~\mathrm{MeV}$, can be reproduced well with $\Lambda_1=2.4~\mathrm{GeV}$. It implies that $D\bar D_1+\mathrm{c.c.}$ occupies sizable component in $Y(4260)$ which is consistent with the current knowledge on the $Y(4260)$ state~\cite{Ding:2008gr,Wang:2013kra,Li:2013yla,Wang:2013cya,Wu:2013onz,Chen:2019mgp,Qin:2016spb,Lu:2017yhl}. And the dozens-of-$\mathrm{MeV}$ width of $D^*\bar{D}^*$ channel is compatible with the calculations in Ref.~\cite{Xue:2017xpu} within the parameter range. The observation that the open charm decay channel $D\bar{D}^*\pi$ accounts for the most of total width of $Y(4260)$ is also reflected in our calculation since the largest two-body channel $D^*\bar{D}^*$ will be entrapped further into the $D\bar{D}^*\pi$ in principle. With the same cutoffs, the total width of $1^{-+}$ $D\bar D_1+\mathrm{c.c.}$ molecule is estimated to be $88.9~\mathrm{MeV}$. Although this value bears a large uncertainty due to the values of the effective coupling constants and the choice of cutoffs $\Lambda_0$ and $\Lambda_1$, the relative ratios among various channels can be regarded as the direct consequence of the molecule structure. In addition to $D^*\bar{D}^*$, $1^{-+}$ state has also large couplings with the $\eta\eta_c$ and $\eta\chi_{c1}$ channels as claimed in Refs.~\cite{Wang:2014wga,Ma:2019hsm}.

It should not be neglected that the partial width of $Y(4260)\to Z_c(3900)\pi$ in our calculation is too small, of order $10^{-5}$ MeV in Table.\ref{tab:widths}. Though no experimental information on this decay channel is available up to now, it may not be that small since the $Y(4260)$ was first discovered in this channel\cite{Ablikim:2013mio}. When we discard the second form factor $f_2(q^2)$, however, the much larger partial width, $1.2~\mathrm{MeV}$, is obtained for the $Z_c \pi$ channel. This value seems to be more acceptable, and it is consistent with the results in Refs.~\cite{Wang:2013cya,Cleven:2013mka,Qin:2016spb} where they also didn't include any suppression on the momentum of exchanged $D^*$ in these channels with $Z_c \pi$ involved. Note that the total widths also get quite large once we discard $f_2(q^2)$. These behavior should be noticed when one would like to compare our results with the experimental data or other estimations from different models. 
\begin{table*}[htbp]
	\centering
	\caption{\label{tab:widths}Partial widths of the $D\bar D_1+\mathrm{c.c.}$ molecule states with quantum numbers $1^{--}$ and $1^{-+}$.  Here $\Lambda_0=1.0$ GeV is fixed. All of decay widths are in unit of $\mathrm{MeV}$ and cutoffs are in unit of $\mathrm{GeV}$. $0$ means that this channel is forbidden by symmetries.}
	\scalebox{1.2}{
		\begin{tabular}{l|*{6}{c}}
			\Xhline{1pt}
			\multirow{3}*{Mode} & \multicolumn{6}{c}{Widths ($\mathrm{MeV}$)} \\
			\Xcline{2-7}{0.4pt}
			& \multicolumn{3}{c}{$1^{--}~Y(4260)$} & \multicolumn{3}{c}{$1^{-+}~\eta_{c1}(4240)$} \\
			\Xcline{2-7}{0.4pt}
			& $\Lambda_1=1.5$ & $\Lambda_1=2.0$& $\Lambda_1=2.4$& $\Lambda_1=1.5$& $\Lambda_1=2.0$& $\Lambda_1=2.4$ \\
			\Xhline{0.8pt}
			$D^*\bar{D}^*$   & 18.1& 26.7 & 31.3& 20.2 & 29.4 &33.5\\
			$\pi Z_c$  		 & $\sim$0& $\sim$0 & $\sim$0& $\sim$0 & $\sim$0 &$\sim$0\\
			$\pi \rho$  	& 0.1& 0.8 & 2.3& 0.06 & 0.4 &1.3\\
			$D \bar{D}$ 	 & 0.02& 0.02 & 0.02& 0 & 0 &0\\
			$\eta\eta_c$ 	 & 0& 0 & 0& 3.9 & 13.9 &22.7\\
			$\eta \chi_{c1}$ & 0& 0 & 0& 4.1 & 11.7 &17.4\\
			$\eta h_c$ 		 & 1.3& 3.9 & 5.8& 0 & 0 &0\\
			$\sigma\chi_{c0}$& 0& 0 & 0& 0.4 & 1.0 &1.3\\
			$\eta J/\psi$    & 0.4& 1.4 & 2.6& 0 & 0&0\\
			$\sigma J/\psi$	 & 0.03& 0.1 & 0.2&0 & 0&0\\
			$\omega \sigma$  & 0.04& 0.3 & 0.8& 0 & 0 &0\\
			$\omega J/\psi$  & 0& 0 & 0& 0.003 & 0.01 &0.02\\
			$\omega \chi_{c0}$  & 0.03& 0.09 & 0.1& 0 & 0 &0\\
			$D \bar{D}^*$    & 0.04& 0.07 & 0.08& 0.04 & 0.06 &0.08\\
			$D \bar{D}^* \pi$    & 1.9& 1.9 & 1.9& 3.0 & 3.0 &3.0\\
			\Xhline{0.8pt}
			Total & 21.9 & 35.3 & 45.1 & 32.1 & 63.2 & 88.9\\
			\Xhline{1pt}
	\end{tabular}}
\end{table*}

\begin{table*}[htbp]
	\centering
	\caption{\label{tab:widths1}Partial widths of the $D\bar D_1+\mathrm{c.c.}$ molecule states with quantum numbers $1^{--}$ and $1^{-+}$. Here the form factor $f_1(q^2)$ is not included. All of decay widths are in unit of $\mathrm{MeV}$ and cutoffs are in unit of $\mathrm{GeV}$. $0$ means that this channel is forbidden by symmetries.}
	\scalebox{1.2}{
		\begin{tabular}{l|*{6}{c}}
			\Xhline{1pt}
			\multirow{3}*{Mode} & \multicolumn{6}{c}{Widths ($\mathrm{MeV}$)} \\
			\Xcline{2-7}{0.4pt}
			& \multicolumn{3}{c}{$1^{--}~Y(4260)$} & \multicolumn{3}{c}{$1^{-+}~\eta_{c1}(4240)$} \\
			\Xcline{2-7}{0.4pt}
			& $\Lambda_0=0.6$ & $\Lambda_0=1.0$& $\Lambda_0=1.4$& $\Lambda_0=0.6$& $\Lambda_0=1.0$& $\Lambda_0=1.4$ \\
			\Xhline{0.8pt}
			$D^*\bar{D}^*$   & 25.3& 51.1 & 86.4& 29.7 & 55.6 &92.5\\
			$\pi Z_c$  		 & 1.2 & 1.2 & 1.2& 1.0 & 1.0 &1.0\\
			$\pi \rho$  	& 7.6& 26.5 & 67.8& 4.4 & 15.6 &38.3\\
			$D \bar{D}$ 	 & 0.006& 0.01 & 0.02& 0 &0&0\\
			$\eta\eta_c$ 	 & 0& 0 & 0& 18.1 & 39.2 &54.8\\
			$\eta \chi_{c1}$ & 0& 0 & 0& 34.1 & 67.4 &98.2\\
			$\eta h_c$ 		 & 13.5 & 26.9 & 40.0& 0 & 0 &0\\
			$\sigma\chi_{c0}$& 0& 0 & 0& 2.4 & 4.4 &6.1\\
			$\eta J/\psi$    & 3.2& 5.6 & 7.1& 0 & 0&0\\
			$\sigma J/\psi$	 & 0.2& 0.4 & 0.6&0 & 0&0\\
			$\omega \sigma$  & 2.5& 6.7 & 12.1& 0 & 0 &0\\
			$\omega J/\psi$  & 0& 0 & 0& 0.01 & 0.03 &0.04\\
			$\omega \chi_{c0}$  & 0.3& 0.5 & 0.7& 0 & 0 &0\\
			$D \bar{D}^*$    & 0.05& 0.1 & 0.1& 0.05 & 0.1 &0.1\\
			$D \bar{D}^* \pi$    & 1.9& 1.9 & 1.9& 3.0 & 3.0 &3.0\\
			\Xhline{0.8pt}
			Total & 55.8 & 120.9 & 218.1 & 151.2 & 296.1 & 392.1\\
			\Xhline{1pt}
	\end{tabular}}
\end{table*}

\section{Summary and Discussion}

In summary, we have used the one boson exchange potential between the $D\bar D_1+\mathrm{c.c.}$, for both $J^{PC}=1^{--}$ and $J^{PC}=1^{-+}$ systems, to investigate if it is possible for them to form bound states. We use the effective Lagrangians, which satisfy the heavy quark symmetry, to describe the interaction between $D$ and $D_1$. We have considered both the vector and $\sigma$ exchange but the latter has little influence. It turns out that with a momentum cutoff $\Lambda\approx 2.25$ GeV, the attractive force between the $D\bar D_1+\mathrm{c.c.}$ with $J^{PC}=1^{--}$ is strong enough to form a bound state with a binding energy of around 60 MeV, corresponding to $Y(4260)$ as a pule $D\bar D_1+\mathrm{c.c.}$ molecule. The C-parity partner of the Y(4260), i.e. the exotic $D\bar D_1+\mathrm{c.c.}$ bound state, is predicted to exist with a mass of about $4240$ MeV, which is consistent with the predictions by lattice QCD and chiral quark model. 

The decay properties of the two possible molecules are investigated by using the Effective Lagrangian Method. For both $1^{--}$ and $1^{-+}$ states, $D^*\bar D^*$ decay channel dominates. The $\eta\eta_c$ and $\eta\chi_{c1}$ channels, absent in the $1^{--}$ case, have sizeable contributions to the $1^{-+}$ molecule decay width, which may help to identify this exotic state experimentally.

Given the $D\bar D_1+\mathrm{c.c.}$ bound states, HQSS indicates the possible existence of other open charm bound states such as $D^*\bar D_1+\mathrm{c.c.}$, $D^{(*)}\bar D_2+\mathrm{c.c.}$ as well as their $D_s$ partners. Recently, Belle Collaboration \cite{Jia:2019gfe} reported a vector charmoniumlike state observed in $e^+e^-\to D_s^+D_{s1}(2536)^- +\mathrm{c.c.}$ with $m=4625.9^{+6.2}_{-6.0}\pm 0.4$ MeV and $\Gamma=49.8^{+13.9}_{-11.5}\pm 4.0$ MeV. This state was considered as a molecule of $D_s^*\bar{D}_{s1}(2536)$ in Ref.~\cite{He:2019csk}. There are also some structures in channels $e^+e^-\to \pi^+\pi^-\psi(3770)$, $\bar DD_1(2400)+\mathrm{c.c.}$ at around $4.4$ GeV \cite{Ablikim:2019faj,Ablikim:2019okg}, which are possible evidences of Y(4360)/Y(4390)/$\psi(4415)$. Further theoretical and experimental studies are needed to identify their structures.

We thank Ying Chen, Yu-Bing Dong, Feng-Kun Guo, Jia-Jun Wu, Mao-Jun Yan, Chang-zheng Yuan and Qiang Zhao for useful discussions and comments. We also give special thanks to Jia-Qi Wang for cross check of parts of this work. This project is supported by NSFC under Grant No. 11621131001 (CRC110 cofunded by DFG and NSFC), Grant No. 11835015 and Grant No. 11947302.

\appendix

\section{Estimation of $g_{\rm{DD}_1\rm V}$\label{sec:gdd1v}}
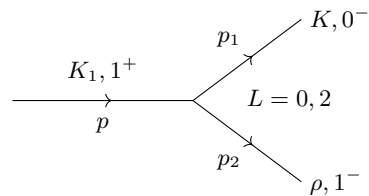
\begin{figure}[h]
	\centering
	
	\begin{tikzpicture}
	\begin{scope}[scale=1.2]
	\draw[electron] (-2,0)--(0,0);
	\draw[electron] (0,0)--(1.2,0.9);
	\draw[electron] (0,0)--(1.2,-0.9);
	
	\node[right] at (1.2,0.9){$K,0^-$};
	
	\node[right] at (1.2,-0.9){$\rho,1^-$};
	\node[right] at (0.5,0){$L=0,2$};
	
	\node[above] at (-1,0.1){$K_1,1^+$};
	
	\node [below] at (-1,-0.1){$p$};
	\node [below] at (0.4,-0.5){$p_2$};
	
	\node [above] at (0.4,0.5){$p_1$};
	\end{scope}
	\end{tikzpicture}
	
	\caption{Feynman diagram for $K_1\to K \rho$ decay. The angular momentum can be $L=0,\ 2$ and we only consider the $L=0$ case.}\label{fig:K1decay}
	\end{figure}

 The partial wave amplitude of the decay $K_1\to K\rho$ for $L=0$ can be expressed as (see e.g. \cite{Zou:2002ar}) 
\begin{equation}
\mathcal M=\sqrt{3/2} g_{K_1K\rho}K_1^{*\mu}(m_1)\rho_\mu(m_2)
\end{equation}
where $K_1^{\mu} (\rho^{\mu})$ is the polarization of $K_1 (\rho)$ and $\sqrt{3/2}$ accounts for the isospin factor, see Fig.~(\ref{fig:K1decay}). 
The decay width reads
\begin{align}
\Gamma=|g_{K_1K\rho}|^2\frac{|\bm{p}_2|}{16\pi m_{K_1}^2}\lra{3+\frac{|\bm{p}_2|^2}{m_\rho^2}}.
\end{align}

In PDG \cite{Tanabashi:2018oca}, there are two different $K_1$ states,
\begin{align}
K_1(1270):\  &m=1272\pm7 \rm{MeV},\notag\\
 &\Gamma=90\pm20 \rm{MeV},\notag\\
&\rm{Br}(K_1(1270)\to K\rho)=(42\pm6)\%,\notag\\
K_1(1400):\  &m=1403\pm7 \rm{MeV},\notag\\
 &\Gamma=174\pm13 \rm{MeV},\notag\\
&\rm{Br}(K_1(1270)\to K\rho)=(3.0\pm3.0)\%,\notag
\end{align}
which lead to 
\begin{align}
g_{K_1(1270)K\rho}&\approx 4.7\ \rm{GeV}\label{eq:gk1270}\\
g_{K_1(1400)K\rho}&\approx 0.75\ \rm{GeV}\label{eq:gk1400}.
\end{align}
So we are faced with a problem, which one to use.

These two mass eigenstates $K_1$ are considered as the mixture of two flavor eigenstates from the axial vector nonets $J^{PC}=1^{++}$ ($^3P_1$) and $J^{PC}=1^{+-}$ ($^1P_1$)\cite{Burakovsky:1997dd, Suzuki:1993yc, Cheng:2003bn, Yang:2010ah, Hatanaka:2008xj, Tayduganov:2011ui, Divotgey:2013jba, Zhang:2017cbi}. It was also explored in \cite{Roca:2005nm, Geng:2006yb, Wang:2019mph} that the $K_1(1270)$ may have a two-pole structure in vector-pseudoscalar scattering since there is always a discrepancy when fitting the experimental data with only one pole\cite{Bowler:1976qe, Daum:1981hb}. Here we ignore the possible two pole structure. 

Following Ref.~\cite{Suzuki:1993yc}, we denote the $^3P_1$ state as $K_a$ and the $^1P_1$ state as $K_b$. The mixing of $K_a$ and $K_b$ is parameterized as
\begin{align}
\left(\begin{array}{c}{\left|K_{a}\right\rangle} \\ {\left|K_{b}\right\rangle}\end{array}\right)=\left(\begin{array}{cc}{\cos \theta_{K}} & {\sin \theta_{K}}\\ {-\sin \theta_{K}} & {\cos \theta_{K}} \end{array}\right)\left(\begin{array}{cc}{\left|K_{1}(1400)\right\rangle}  \\ {\left|K_{1}(1270)\right\rangle}\end{array}\right)
\end{align}
and the mixing angle is determined to be around $33^{\circ}$ or $58^{\circ}$ \cite{Burakovsky:1997dd, Suzuki:1993yc, Cheng:2003bn, Yang:2010ah, Hatanaka:2008xj, Tayduganov:2011ui, Divotgey:2013jba, Zhang:2017cbi}.

On the other hand, as mentioned in the Introduction, the decay width of $D_1(2420)$ is much smaller than that of $D_1(2430)$. This phenomenon can be explained in the heavy quark limit, where the spin of heavy quark, $\bm s_h$, is decoupled with the angular momentum of the light quark, $\bm s_l=\bm s_q+\bm L$ with $\bm s_q$ the spin of light quark and $\bm L$ the orbital angular momentum. Note that $D_1$ with $s_l=1/2$ can decay into $D^*\pi$ in S-wave while such decay for $D_1$ with $s_l=3/2$ can only happen in $D$-wave, which has a smaller width. Therefore, $D_1(2420)$ is regarded as the state $\ket{s_h=1/2, s_l=3/2,j=1,m}$, which can be decomposed into the $L-S$ basis,
\begin{align}
\ket{j=1,m}&=\sqrt{\frac13}\ket{l=1,s=1,j=1,m}\notag\\
&-\sqrt{\frac{2}{3}}\ket{l=1,s=0,j=1,m}.
\end{align}
In other words, 
\begin{align}
\ket{D_1(2420)}=\sqrt{\frac13}\ket{D_a}-\sqrt{\frac{2}{3}}\ket{D_b}
\end{align}
where $D_a$ and $D_b$ correspond to $K_a$ and $K_b$, respectively. Therefore, the $K_1$ state corresponding to $\ket{D_1(2420)}$, denoted by $\ket{\tilde K_1}$, is related to the two mass eigenstates via
\begin{align}
\ket{\tilde K_1}&=\sqrt{\frac13}\ket{K_a}-\sqrt{\frac{2}{3}}\ket{K_b}\notag\\
&=\lra{\sqrt{\frac13}\cos \theta_{K}+\sqrt{\frac23}\sin \theta_{K}}\left|K_{1}(1400)\right\rangle\notag\\
&+\lra{\sqrt{\frac13}\sin \theta_{K}-\sqrt{\frac23}\cos \theta_{K}}\left|K_{1}(1270)\right\rangle.\label{eq:gk2420}
\end{align}
Substituting $\theta_K=33^{\circ}\ {\rm or}\ 58^{\circ}$ into Eq.\ref{eq:gk2420} and together with Eqs.~(\ref{eq:gk1270},\ref{eq:gk1400}) we get
\begin{align}
g_{\tilde K_1KV}
\approx\mp1.0\ \rm{GeV}.
\end{align}
Now $g_{\rm {DD}_1(2420)\rm V}$ can be estimated via Eq.~(\ref{eq:gDK})
\begin{align}
g_{\rm {DD}_1(2420)\rm V}\approx 0.9\frac{\sqrt{m_{D_1}m_D}}{\sqrt{m_{\tilde K_1}m_K}}g_{\tilde K_1KV}\approx \mp 2.3\ \rm{GeV}
\end{align}
where we have used $m_{\tilde K_1}=1390$ MeV as the approximation of the expectation value of $\tilde K_1$'s mass. Actually, the sign of $g_{\rm {DD}_1(2420)\rm V}$ does not matter in our calculation so we just use the positive one.

\bibliography{ref}

\end{document}